\documentstyle[12pt]{article}
\begin{document}
\baselineskip=15.5pt
\begin{titlepage}

\begin{flushright}
IC/2002/14\\
hep-th/0203145
\end{flushright}
\vspace{10 mm}

\begin{center}
{\Large Variable-Speed-of-Light Cosmology and\\ 
Second Law of Thermodynamics}

\vspace{5mm}

\end{center}

\vspace{5 mm}

\begin{center}
{\large Donam Youm\footnote{E-mail: youmd@ictp.trieste.it}}

\vspace{3mm}

ICTP, Strada Costiera 11, 34014 Trieste, Italy

\end{center}

\vspace{1cm}

\begin{center}
{\large Abstract}
\end{center}

\noindent

We examine whether the cosmologies with varying speed of light (VSL) are 
compatible with the second law of thermodynamics.  We find that the VSL 
cosmology with varying fundamental constant is severely constrained by the 
second law of thermodynamics, whereas the bimetric cosmological models are 
less constrained.

\vspace{1cm}
\begin{flushleft}
March, 2002
\end{flushleft}
\end{titlepage}
\newpage

\section{Introduction}

Variable-Speed-of-Light (VSL) cosmological models were proposed 
\cite{mof1,am} as an alternative to inflation \cite{gut,lin,als} for solving 
the initial value problems in the standard Big Bang model.  It is assumed 
in the VSL models that the speed of light initially took a larger value 
and then decreased to the present day value during an early period of 
cosmic evolution.  VSL models have attracted some attention, because not 
only various cosmological problems that are solved by the inflationary 
models but also the cosmological constant problem can be solved 
\cite{mof1,am,bar1,bar2,bar3,mof2,cm1,bar4,cm2,blm,cm3,cm4} by VSL models.  
Furthermore, the recent study of quasar absorption line spectra in comparison 
with laboratory spectra shows that the fine structure constant $\alpha=
e^2/(4\pi\hbar c)$ varies over cosmological time scales \cite{wfc,mur,web}, 
indicating that the speed of light may indeed vary with time.  Also, it has 
been shown \cite{kal1,kir,kal2,chu,ale,ish,ckr,csa} that brane world 
models, which have been in vogue recently, manifest the Lorentz violation, 
which is a necessary requirement for the VSL models.  (Cf. The recent work 
\cite{bcf} studies the experimental limits which are permitted for the 
graviton's speed in the brane world scenarios.)

It is the purpose of this paper to examine the compatibility of the VSL models 
with the second law of thermodynamics.  (Cf. The previous related work can be 
found in Ref. \cite{cjp}.)  Recently, there has been active 
interest in holographic principle in cosmology, after the initial 
work by Fischler and Susskind (FS) \cite{fs}.  The cosmological holographic 
bound originally formulated by FS had a problem of being violated by the 
closed Friedmann-Robertson-Walker (FRW) universe.  Later works 
attempted to circumvent such a problem through various modifications.  
In particular, it was proposed in Refs. \cite{el,ram} that the FS 
holographic bound has to be replaced by the generalized second law of 
thermodynamics.  The generalized second law states that the total entropy $S$ 
of the universe should not decrease with time during the cosmological 
evolution: $dS\geq 0$.  In order to be compatible with the holographic 
principle, the VSL cosmological models therefore have to obey the generalized 
second law of thermodynamics.  In section 2, we consider the original VSL 
model, where the speed of light $c$ in the action is just assumed to vary with 
time.  In section 3, we consider the bimetric cosmology of Clayton and 
Moffat.  

\section{VSL Cosmology with Varying Fundamental Constant}

First, we consider the original VSL cosmology \cite{mof1,am}, in which a  
fundamental constant $c$ of the nature is just assumed to vary with time 
during the early period of the cosmic evolution and thereby the Lorentz 
symmetry is explicitly broken.  In such VSL theories, it is postulated 
that there exists a preferred Lorentz frame in which laws of physics simplify 
with the action taking a standard form with a constant $c$ replaced by a 
field $c(x^{\mu})$, {\it the principle of minimal coupling}.  Namely, the 
action in the preferred frame takes the form
\begin{equation}
S=\int d^4x\left[\sqrt{-g}\left\{{c^4\over{16\pi G}}\left({\cal R}-2\Lambda
\right)+{\cal L}\right\}+{\cal L}_c\right],
\label{cosact}
\end{equation}
where ${\cal L}_c$ controls the dynamics of $c$ and ${\cal L}$ is the action 
for the fields in the universe.  It is required that ${\cal L}_c$ should be 
explicitly independent of the other fields, including metric, so that the 
principle of minimal coupling continues to hold for the equations of motion.  

The general metric ansatz for a 4-dimensional homogeneous and isotropic 
universe is given by the following Robertson-Walker metric:
\begin{equation}
g_{\mu\nu}dx^{\mu}dx^{\nu}=-c^2dt^2+a^2\gamma_{ij}dx^idx^j,
\label{rwmet}
\end{equation}
with the time-varying $c$.  Here, $a(t)$ is the cosmic scale factor and 
$\gamma_{ij}(x^k)$ is given by
\begin{equation}
\gamma_{ij}dx^idx^j=\left(1+\textstyle{k\over 4}\delta_{mn}x^mx^n\right)^{-2}
\delta_{ij}dx^idx^j={{dr^2}\over{1-kr^2}}+r^2d\Omega^2_2,
\label{gammet}
\end{equation}
where $k=-1,0,1$ respectively for the open, flat and closed universes.  

With an assumption of the principle of minimal coupling, the Einstein 
equations with the metric ansatz (\ref{rwmet}) lead to the Friedmann equations:
\begin{equation}
\left({\dot{a}\over a}\right)^2={{8\pi G}\over 3}\rho+{c^2\over 3}\Lambda
-{{kc^2}\over a^2},
\label{frd1}
\end{equation}
\begin{equation}
{\ddot{a}\over a}=-{{4\pi G}\over 3}\left(\rho+3{p\over c^2}\right)
+{c^2\over 3}\Lambda,
\label{frd2}
\end{equation}
where the overdot denotes derivative w.r.t. $t$.
From the Friedmann equations, we obtain the following generalized conservation 
equation:
\begin{equation}
\dot{\rho}+3\left(\rho+{p\over c^2}\right){\dot{a}\over a}
={{3kc\dot{c}}\over{4\pi Ga^2}}-{{c\dot{c}}\over{4\pi G}}\Lambda.
\label{gencnsveq}
\end{equation}

We now study thermodynamics of the VSL cosmology.  We assume that the 
universe satisfies the first law of thermodynamics.  When applied to the 
comoving volume element of unit coordinate volume and physical volume 
$v=a^3$, the first law of thermodynamics takes the form:
\begin{equation}
Tds=d(\rho c^2v)+pdv,
\label{1stthlaw}
\end{equation}
where $s=s(v,T)$ is the entropy density of the universe at temperature $T$ 
within the volume $v$, and $\rho=\rho(T)$ and $p=p(T)$ are the mass density 
and the pressure of matter in the universe.  In this paper, we assume that 
$c$ is a function of $a$, just as in Ref. \cite{bar3}.  Since $v=a^3$, we 
have $c^{\prime}(a)=3a^2dc/dv=3v^{2/3}dc/dv$, where prime denotes derivative 
w.r.t. $a$.  Then, Eq. (\ref{1stthlaw}) can be rewritten as
\begin{eqnarray}
Tds&=&c^2vd\rho+\left(\rho c^2+p+2\rho cv{{dc}\over {dv}}\right)dv
\cr
&=&c^2vd\rho+\left(\rho c^2+p+\textstyle{2\over 3}\rho cc^{\prime}v^{1/3}
\right)dv.
\label{1stlaw}
\end{eqnarray}
So, partial derivatives of $s(v,T)$ are given by
\begin{equation}
{{\partial s(v,T)}\over{\partial v}}={1\over T}\left(\rho c^2+p+
\textstyle{2\over 3}\rho cc^{\prime}v^{1/3}\right),
\label{sdv}
\end{equation}
\begin{equation}
{{\partial s(v,T)}\over{\partial T}}={{c^2v}\over T}{{d\rho}\over{dT}}.
\label{sdt}
\end{equation}
From the integrability condition $\partial^2s/(\partial v\partial T)=
\partial^2s/(\partial T\partial v)$, we obtain
\begin{equation}
{{dp}\over{dT}}={1\over T}\left(\rho c^2+p+\textstyle{2\over 3}\rho c
c^{\prime}v^{1/3}\right)={1\over T}{d\over{dv}}\left[(\rho c^2+p)v\right].
\label{intcond}
\end{equation}
Making use of this equation, we can put Eq. (\ref{1stthlaw}) into the 
following form:
\begin{equation}
ds=d\left[{v\over T}(\rho c^2+p)\right]-{{2v^2}\over T^2}\rho c{{dc}\over
{dv}}dT,
\label{1stlaw2}
\end{equation}
from which we see that the usual Euler's relation $s={v\over T}(\rho c^2+p)$ 
does not hold for the VSL theories with varying fundamental constant.  

To obtain the time derivative of $s$, we express the conservation 
equation (\ref{gencnsveq}) into the following form, making use of Eq. 
(\ref{intcond}):
\begin{equation}
{d\over{dt}}\left[{v\over T}(\rho c^2+p)\right]=\left(2\rho{\dot{c}\over
c}-{{c\dot{c}}\over{4\pi G}}\Lambda+{{3kc\dot{c}}\over{4\pi Ga^2}}\right)
{{c^2v}\over T}+2\rho c{{dc}\over{dv}}v^2{\dot{T}\over T^2}.
\label{auxeq}
\end{equation}
Eq. (\ref{1stlaw2}) along with Eq. (\ref{auxeq}) yields
\footnote{The previous related work \cite{cjp} does not take into account 
the effects on $\dot{s}$ of the modification of the usual Euler's relation 
due to the time varying $c$ and the nonzero cosmological constant $\Lambda$.}
\begin{equation}
\dot{s}=\left(2\rho-{c^2\over{4\pi G}}\Lambda+{{3kc^2}\over{4\pi Ga^2}}
\right){{c\dot{c}a^3}\over T}.
\label{sdotvf}
\end{equation}
This equation would also have been obtained directly from Eqs. 
(\ref{gencnsveq}) and (\ref{1stthlaw}).  

Since $\dot{c}<0$ for the VSL cosmology with varying fundamental constant, 
the terms in the parenthesis of Eq. (\ref{sdotvf}) have to be non-positive 
in order to be compatible with the second law of thermodynamics $dS\geq 0$:
\begin{equation}
2\rho-{c^2\over{4\pi G}}\Lambda+{{3kc^2}\over{4\pi Ga^2}}\leq 0.
\label{vslfp2nd}
\end{equation}  
When the cosmological constant is non-positive ($\Lambda\leq 0$), this 
condition can never be satisfied by the flat ($k=0$) and the closed ($k=1$) 
universes.  Although it can be satisfied by the open universe ($k=-1$), 
the condition is very restrictive about the possible type of matter in the 
universe and the time variation of $c$.  On the other hand, when the 
cosmological constant is positive ($\Lambda>0$) and is large enough, the 
restriction becomes less severe.   We now examine the explicit condition 
under which the constraint (\ref{vslfp2nd}) can be satisfied.  Assuming 
that $c$ varies gradually like $c(t)=c_0a^n$ and the equation of state of 
the form $p=(\gamma-1)\rho c^2$, by solving Eq. (\ref{gencnsveq}) one obtains
\begin{equation}
\rho={B\over a^{3\gamma}}+{{3kc^2_0na^{2(n-1)}}\over{4kG(3\gamma+2n-2)}}-
{{\Lambda nc^2_0a^{2n}}\over{4\pi G(3\gamma+2n)}},
\label{rhotoa}
\end{equation}
where $B$ is a positive constant.  So, the constraint (\ref{vslfp2nd}) 
reduces to
\begin{equation}
{{2B}\over a^{3\gamma+2n}}+{{3(3\gamma+4n-2)kc^2_0}\over{4\pi G(3\gamma+2n-2)
a^2}}\leq {{(3\gamma+4n)c^2_0\Lambda}\over{4\pi G(3\gamma+2n)}}.
\label{cstrant}
\end{equation}
First of all, we see from this that for the flat universe ($k=0$) the 
constraint is always violated for some values of $a$.  [When $3\gamma+2n=0$, 
the last term on the RHS of Eq. (\ref{rhotoa}) behaves with $a$ as $\sim 
a^{-3\gamma}\ln a$ and therefore the RHS of Eq. (\ref{cstrant}) behaves 
with $a$ as $\sim\ln a$.]  In order for Eq. (\ref{cstrant}) to be satisfied 
for any values of $a$, the LHS has to have the maximum value and its value 
has to be not greater than the RHS.  From such condition, we obtain the 
following constraint on the constants $n$ and $\gamma$:
\begin{equation}
k(3\gamma+4n-2)>0,\ \ \ \ \ \ \ \ \ \ 
(3\gamma+2n)(3\gamma+2n-2)<0,
\label{cnstcnstrnt}
\end{equation}
along with the following lower limit on the cosmological constant:
\begin{equation}
\Lambda\geq {{3(3\gamma+4n-2)k}\over{3\gamma+4n}}\left[-{{3(3\gamma+4n-2)
kc^2_0}\over{4\pi(3\gamma+2n)(3\gamma+2n-2)GB}}\right]^{2\over{3\gamma+2n-2}}.
\label{cscnst}
\end{equation}
One can always choose the integration constant $B$ in such a way that 
the bound (\ref{cscnst}) is compatible with the observed value of 
$\Lambda$.  However, the constraint (\ref{cnstcnstrnt}) severely 
restricts the allowed value of $\gamma$.  For the closed universe 
($k=1$), Eq. (\ref{cnstcnstrnt}) implies $\gamma<2/3$ and therefore 
the radiation-dominated universe ($\gamma=4/3$) and the 
matter-dominated universe ($\gamma=1$) are not allowed.  As for the 
open universe ($k=-1$), Eq. (\ref{cnstcnstrnt}) leads to the less 
severe constraint $\gamma>-2/3$, which allows the $\gamma=4/3,1$ 
cases.  

\section{Scalar-Tensor Bimetric Cosmology}

In this section, we consider the bimetric VSL model, proposed by Clayton and 
Moffat \cite{cm1}.  The bimetric models achieve time-variable speed of light 
in a diffeomorphism invariant manner and without explicitly breaking the 
Lorentz symmetry by introducing bimetric structure into spacetime.  (Cf. 
It is recently found out in Ref. \cite{moff} that the fine-structure constant 
$\alpha=e^2/(4\pi\hbar c)$ in the bimetric models is constant in spacetime 
although the speed of light varies with time, due to the compensating 
time-variation of the electric charge.)  It is usually assumed in the 
bimetric models that graviton and the biscalar (or the bivector) propagate 
on the geometry described by the ``gravity metric'', whereas all the matter 
fields (including photons) propagate on the geometry described by the 
``matter metric''.  In the case of the scalar-tensor bimetric model, the 
gravity metric $g_{\mu\nu}$ and the matter metric $\hat{g}_{\mu\nu}$ are 
related by the biscalar field $\Phi$ as
\begin{equation}
\hat{g}_{\mu\nu}=g_{\mu\nu}-B\partial_{\mu}\Phi\partial_{\nu}\Phi,
\label{metrel}
\end{equation}
where a dimensionless constant $B$ is assumed to be positive.  Since these two 
metrics are nonconformally related, a photon and a graviton propagate at 
different speeds.  The action for the scalar-tensor bimetric model has the form
\begin{equation}
S=\int d^4x\sqrt{-g}\left[{c^4\over{16\pi G}}\left({\cal R}-2\Lambda\right)
+{\cal L}_{\Phi}\right]+\int d^4x\sqrt{-\hat{g}}{\cal L}_{\rm mat},
\label{bimact}
\end{equation}
where ${\cal L}_{\rm mat}$ is the Lagrangian density for matter fields and 
the Lagrangian density ${\cal L}_{\Phi}$ for the biscalar is given by
\begin{equation}
{\cal L}_{\Phi}=-{1\over 2}g^{\mu\nu}\partial_{\mu}\Phi\partial_{\nu}\Phi
-V(\Phi).
\label{bisact}
\end{equation}
The gravity metric for the universe has the form
\begin{equation}
g_{\mu\nu}dx^{\mu}dx^{\nu}=-c^2dt^2+a^2\gamma_{ij}dx^idx^j,
\label{gravcsmet}
\end{equation}
with constant speed of graviton $c_{\rm grav}=c$ and $\gamma_{ij}$ given by 
Eq. (\ref{gammet}).  Due to the requirement of homogeneity and isotropy 
of the universe, the biscalar field $\Phi$ is independent of the spatial 
coordinates $x^i$.  According to Eq. (\ref{metrel}), the matter metric is 
therefore given by
\begin{equation}
\hat{g}_{\mu\nu}dx^{\mu}dx^{\nu}=-(c^2+B\dot{\Phi}^2)dt^2
+a^2\gamma_{ij}dx^idx^j,
\label{matcsmet}
\end{equation}
where the overdot stands for derivative w.r.t. $t$.  So, the speed of 
photon $c_{\rm ph}=c\sqrt{1+B\dot{\Phi}^2/c^2}\equiv c\sqrt{I}$ 
varies with $t$, taking larger value than $c_{\rm grav}=c$ while 
$\dot{\Phi}\neq 0$.  

In obtaining the energy-momentum tensor for the purpose of deriving the 
Einstein's equations, one has to keep in mind that matter fields and 
biscalar are coupled to different metrics.  Since $\hat{g}_{\mu\nu}$ is 
the physical metric for the matter fields, the energy-momentum tensor for 
the matter fields are defined in terms of $\hat{g}_{\mu\nu}$:
\begin{equation}
\hat{T}^{\mu\nu}\equiv{2\over\sqrt{-\hat{g}}}{{\delta(\sqrt{-\hat{g}}
{\cal L}_{\rm mat})}\over{\delta\hat{g}_{\mu\nu}}}=\left(\hat{\rho}
c^2_{\rm ph}+\hat{p}\right)\hat{U}^{\mu}\hat{U}^{\nu}+\hat{p}\hat{g}^{\mu\nu},
\label{emmat}
\end{equation}
where $\hat{\rho}$ and $\hat{p}$ are the mass density and the pressure of the 
matter fields and $\hat{U}^{\mu}$ is the four-velocity of matter perfect fluid 
normalized as $\hat{g}_{\mu\nu}\hat{U}^{\mu}\hat{U}^{\nu}=-1$.  Since the 
nonzero component of the four-velocity vector in the comoving coordinates 
is $\hat{U}^t=1/c_{\rm ph}$, the nonzero components of the energy-momentum 
tensor for the matter fields are
\begin{equation}
\hat{T}^{tt}=\hat{\rho},\ \ \ \ \ \ \ 
\hat{T}^{ij}={\hat{p}\over a^2}\gamma^{ij}.
\label{emmatcomp}
\end{equation}
On the other hand, since the biscalar field is coupled to the gravity 
metric $g_{\mu\nu}$, its energy-momentum tensor is defined in terms of 
$g_{\mu\nu}$:
\begin{eqnarray}
T^{\mu\nu}_{\Phi}&\equiv&{2\over\sqrt{-g}}{{\delta(\sqrt{-g}{\cal L}_{\Phi})}
\over{\delta g_{\mu\nu}}}=g^{\mu\alpha}g^{\nu\beta}\partial_{\alpha}\Phi
\partial_{\beta}\Phi-\textstyle{1\over 2}g^{\mu\nu}\partial_{\alpha}\Phi
\partial^{\alpha}\Phi-V(\Phi)g^{\mu\nu}
\cr
&=&\left(\rho_{\Phi}c^2+p_{\Phi}\right)U^{\mu}
U^{\nu}+p_{\Phi}g^{\mu\nu},
\label{bistens}
\end{eqnarray}
where the four-velocity $U^{\mu}$ for the biscalar is normalized as 
$g_{\mu\nu}U^{\mu}U^{\nu}=1$, so its nonzero component is $U^t=1/c$.  The 
mass density and the pressure of the biscalar field are therefore
\begin{equation}
\rho_{\Phi}=\left({1\over 2}{\dot{\Phi}^2\over c^2}+V\right){1\over c^2}, 
\ \ \ \ \ \ \ \ 
p_{\Phi}={1\over 2}{\dot{\Phi}^2\over c^2}-V.
\label{mssprsbs}
\end{equation}

Taking the variation of the action $S$ w.r.t. the metric, we obtain the  
Einstein's equations
\begin{equation}
{\cal G}^{\mu\nu}+\Lambda g^{\mu\nu}={{8\pi G}\over c^4}{\cal T}^{\mu\nu},
\label{eineqs}
\end{equation}
where ${\cal G}_{\mu\nu}$ is the Einstein tensor for $g_{\mu\nu}$ and the 
energy-momentum tensor ${\cal T}_{\mu\nu}$ has the form
\begin{equation}
{\cal T}^{\mu\nu}=T^{\mu\nu}_{\Phi}+\hat{T}^{\mu\nu}{\sqrt{-\hat{g}}\over
\sqrt{-g}}.
\label{totemten}
\end{equation}
The nonzero components of ${\cal T}^{\mu\nu}$ are
\begin{equation}
{\cal T}^{tt}=\rho_{\Phi}+\hat{\rho}{c_{\rm ph}\over c},\ \ \ \ \ \ 
{\cal T}^{ij}=\left(p_{\Phi}+\hat{p}{c_{\rm ph}\over c}\right)
{1\over a^2}\gamma^{ij}.
\label{nontens}
\end{equation}
The Einstein's equations lead to the Friedmann equations
\begin{equation}
\left({\dot{a}\over a}\right)^2={{8\pi G}\over 3}\rho+{c^2\over 3}\Lambda
-{{kc^2}\over a^2},
\label{fredeq1}
\end{equation}
\begin{equation}
{\ddot{a}\over a}=-{{4\pi G}\over 3}\left(\rho+3{p\over c^2}\right)+
{c^2\over 3}\Lambda,
\label{fredeq2}
\end{equation}
where
\begin{equation}
\rho\equiv\rho_{\Phi}+\hat{\rho}{c_{\rm ph}\over c},\ \ \ \ \ \ \ 
p\equiv p_{\Phi}+\hat{p}{c_{\rm ph}\over c}.
\label{rhopdefs}
\end{equation}
From these Friedmann equations, we obtain the following conservation equation
\begin{equation}
\dot{\rho}+3\left(\rho+{p\over c^2}\right){\dot{a}\over a}=0.
\label{conseq}
\end{equation}

The factors of $c_{\rm ph}/c_{\rm grav}=c_{\rm ph}/c$ in Eq. (\ref{rhopdefs}) 
can be understood from the fact that definition of the energy-momentum tensor 
depends on the choice of metric.  Generally, the following two 
energy-momentum tensors, associated with the same Lagrangian density $L$ but 
defined w.r.t. the two different metrics $g_{\mu\nu}$ and $\hat{g}_{\mu\nu}$,
\begin{equation}
T^{\mu\nu}_{\rm grav}\equiv {2\over\sqrt{-g}}{{\delta L}\over
{\delta g_{\mu\nu}}},\ \ \ \ \ \ \ \ \ \ \ \ \ \ \ 
T^{\mu\nu}_{\rm ph}\equiv{2\over\sqrt{-\hat{g}}}{{\delta L}\over
{\delta\hat{g}_{\mu\nu}}},
\label{twoemten}
\end{equation}
are related to each other as
\begin{equation}
T^{\mu\nu}_{\rm grav}={\sqrt{-\hat{g}}\over\sqrt{-g}}T^{\mu\nu}_{\rm ph}=
{c_{\rm ph}\over c_{\rm grav}}T^{\mu\nu}_{\rm ph}.
\label{emtenrel}
\end{equation}
Here, the subscripts `grav' and `ph' signify that the quantity under 
consideration is defined w.r.t. the gravity and the matter metrics, 
respectively.  The mass densities and the pressures in the two different 
definitions are defined by
\begin{eqnarray}
T^{\mu\nu}_{\rm grav}&=&\left(\rho_{\rm grav}c^2_{\rm grav}+p_{\rm grav}\right)
U^{\mu}U^{\nu}+p_{\rm grav}g^{\mu\nu},
\cr
T^{\mu\nu}_{\rm ph}&=&\left(\rho_{\rm ph}c^2_{\rm ph}+p_{\rm ph}\right)
\hat{U}^{\mu}\hat{U}^{\nu}+p_{\rm ph}\hat{g}^{\mu\nu},
\label{mssprsdefs}
\end{eqnarray}
where the four-velocities are normalized as $g_{\mu\nu}U^{\mu}U^{\nu}=-1$ and 
$\hat{g}_{\mu\nu}\hat{U}^{\mu}\hat{U}^{\nu}=-1$.  From Eqs. (\ref{emtenrel}) 
and (\ref{mssprsdefs}), we see that the mass densities and the pressures in 
the two different definitions are related as
\begin{equation}
\rho_{\rm grav}={c_{\rm ph}\over c_{\rm grav}}\rho_{\rm ph},
\ \ \ \ \ \ \ \ \ \ \ \ 
p_{\rm grav}={c_{\rm ph}\over c_{\rm grav}}p_{\rm ph}.
\label{mssprsrel}
\end{equation}
The factor of $c_{\rm grav}/c_{\rm ph}$, multiplying the matter fields mass 
density and pressure in Eq. (\ref{rhopdefs}), arose due to the fact that 
the matter fields mass density and pressure, which are defined w.r.t. the 
matter metric, has to be transformed to the `gravity metric' quantities, 
since the Friedmann equations are defined w.r.t. the gravity metric.  
(Cf. In the Einstein's equations (\ref{eineqs}), the total energy-momentum 
tensor ${\cal T}^{\mu\nu}$ is defined w.r.t. the gravity metric, i.e., 
${\cal T}^{\mu\nu}\equiv{2\over\sqrt{-g}}{{\delta L}\over{\delta 
g_{\mu\nu}}}$ where Lagrangian density $L$ is the sum of the matter fields 
Lagrangian density $L_{\rm mat}=\sqrt{-\hat{g}}{\cal L}_{\rm mat}$ and the 
biscalar field Lagrangian density $L_{\Phi}=\sqrt{-g}{\cal L}_{\Phi}$.)  

We now study thermodynamics of the bimetric VSL cosmology.  Since we are 
considering the frame associated with the gravity metric of the form 
(\ref{gravcsmet}), the physical quantities in the thermodynamic laws should 
be `gravity metric' quantities.  The first law of thermodynamics, applied to 
the comoving volume element of unit coordinate volume and physical volume 
$v=a^3$, therefore takes the form
\begin{equation}
Tds=d(\rho c^2v)+pdv,
\label{1stlw1}
\end{equation}
where $\rho$ and $p$ are given by Eq. (\ref{rhopdefs}).  Note, the square 
of the speed of graviton $c_{\rm grav}=c$ multiplies $\rho$ to yield the 
energy density, because we are considering the frame associated with the 
gravity metric.   From Eqs. (\ref{conseq}) and (\ref{1stlw1}), we obtain 
$ds/dt=0$, namely the total entropy $S=s\int dx^3\sqrt{-\gamma}$ of the 
universe remains constant during the cosmic evolution.   Therefore, unlike 
the VSL cosmology with varying fundamental constant, considered in the 
previous section, the second law of thermodynamics $dS\geq 0$ is always 
obeyed regardless of values of $k$. 

We comment on thermodynamics associated with the matter fields, only.  
Since it is assumed that the matter field action is constructed out of 
$\hat{g}_{\mu\nu}$, the equations of motion of the matter fields imply the 
conservation law for the matter fields energy-momentum tensor \cite{cm2}:
\begin{equation}
\hat{\nabla}_{\mu}\hat{T}^{\mu\nu}=0,
\label{matencons}
\end{equation}
where $\hat{\nabla}_{\mu}$ is the covariant derivative defined w.r.t. 
$\hat{g}_{\mu\nu}$.  The $t$-component $\hat{\nabla}_{\mu}\hat{T}^{\mu}_{\ 
\ t}=0$ of the conservation equation, where $\hat{T}^{\mu}_{\ \nu}\equiv 
\hat{T}^{\mu\rho}\hat{g}_{\rho\nu}$, takes the following form:
\begin{equation}
\dot{\hat{\rho}}+3\left(\hat{\rho}+{\hat{p}\over c^2_{\rm ph}}\right)
{\dot{a}\over a}=-2\hat{\rho}{\dot{c}_{\rm ph}\over c_{\rm ph}}.
\label{conseqbs}
\end{equation}
From this equation we see that matter fields are created while the speed 
of photon decreases with time to the present day value $c$.  In the frame 
associated with the matter metric, the first law of thermodynamics takes 
the form:
\begin{equation}
Tds_{\rm mat}=d(\hat{\rho}c^2_{\rm ph}v)+\hat{p}dv,
\label{fstlw}
\end{equation}
where $s_{\rm mat}$ is the entropy density associated with the matter fields.  
From Eqs. (\ref{conseqbs}) and (\ref{fstlw}), we see that $\dot{s}_{\rm mat}
=0$, namely that the entropy for the matter fields remain constant despite 
that the matter fields are created while $\dot{c}_{\rm ph}<0$.  This apparent 
paradox can be understood from the fact that in the matter metric frame the 
past light cone contracts and thereby less information is collected by the 
observer.  Next, we consider the frame associated with the gravity metric.  
In this frame, the mass density and the pressure of the matter fields are 
transformed to $\hat{\rho}c_{\rm ph}/c$ and $\hat{p}c_{\rm ph}/c$, and the 
square of the speed of graviton $c$ should be multiplied to the mass density 
for obtaining the energy density.  So, the first law of thermodynamics takes 
the form:
\begin{equation}
Tds_{\rm mat}=d(\hat{\rho}cc_{\rm ph}v)+{c_{\rm ph}\over c}pdv.
\label{sndlw2}
\end{equation}
From Eqs. (\ref{conseqbs}) and (\ref{sndlw2}), we have
\begin{equation}
\dot{s}_{\rm mat}=\left(3{{c^2_{\rm ph}-c^2}\over{c^2c^2_{\rm ph}}}\hat{p}
{\dot{a}\over a}-\hat{\rho}{\dot{c}_{\rm ph}\over c_{\rm ph}}\right)
{{cc_{\rm ph}a^3}\over T}.
\label{sdot2}
\end{equation}
So, in the gravity frame, the entropy for the matter fields varies with time  
while $\dot{c}_{\rm mat}\neq 0$.  This time variation of the entropy of the 
matter fields is due to the exchange of entropy with the biscalar sector, 
since we have seen that the total entropy density $s$ remains constant in 
the gravity metric frame.  

In the above, we considered the equations in the comoving frame for the 
gravity metric $g_{\mu\nu}$.  Since all the matter fields in the universe 
are coupled to the matter metric $\hat{g}_{\mu\nu}$, it would be more natural 
to consider the comoving frame for the matter metric in studying the expansion 
of the universe.  By defining the cosmic time $\tau$ for the matter metric 
in the following way
\begin{equation}
d\tau^2\equiv(1+B\dot{\Phi}^2/c^2)dt^2,
\label{cstmmm}
\end{equation} 
we can bring the matter metric into the following standard comoving frame 
form for the Robertson-Walker metric:
\begin{equation}
\hat{g}_{\mu\nu}dx^{\mu}dx^{\nu}=-c^2d\tau^2+a^2(\tau)\gamma_{ij}dx^idx^j.
\label{mmrw}
\end{equation}
In this new frame, the gravity metric (\ref{gravcsmet}) takes the form
\begin{equation}
g_{\mu\nu}dx^{\mu}dx^{\nu}=-(c^2-B\dot{\Phi}^2)d\tau^2+a^2(\tau)\gamma_{ij}
dx^idx^j,
\label{mmfgm}
\end{equation}
where the overdot from now on stands for derivative w.r.t. $\tau$.  So, in 
this new frame, a photon travels with a constant speed $c_{\rm ph}=c$ and a 
graviton travels with a time-variable speed $c_{\rm grav}=\sqrt{c^2-B
\dot{\Phi}^2}=c/\sqrt{I}$, taking smaller value than $c$ while $\dot{\Phi}
\neq 0$.  Note, $I=1/(1-B\dot{\Phi}^2/c^2)$ when the overdot stands for 
derivative w.r.t. $\tau$.  

In obtaining the Friedmann equations in the new frame, we do not just apply 
the change of time coordinate (\ref{cstmmm}) in the Friedmann equations 
(\ref{fredeq1}) and (\ref{fredeq2}) in the old frame, unlike the previous 
works on bimetric VSL cosmology.  The reason is that the definitions for 
the mass density and the pressure depend on the choice of time coordinate.  
We consider the Einstein's equations (\ref{eineqs}) with ${\cal G}_{\mu\nu}$ 
now being the Einstein tensor for the gravity metric given by Eq. 
(\ref{mmfgm}).  The energy-momentum tensor ${\cal T}_{\mu\nu}$ is still 
given by Eq. (\ref{totemten}) but now with
\begin{equation}
\hat{T}^{\mu\nu}=\left(\hat{\rho}c^2+\hat{p}\right)\hat{U}^{\mu}\hat{U}^{\nu}
+\hat{p}\hat{g}^{\mu\nu},
\label{matem2}
\end{equation}
\begin{equation}
T^{\mu\nu}_{\Phi}=\left(\rho_{\Phi}c^2_{\rm grav}+p_{\Phi}\right)U^{\mu}
U^{\nu}+p_{\Phi}g^{\mu\nu},
\label{phem2}
\end{equation}
where the nonzero components of the four-velocities are $\hat{U}^t=1/c$ 
and $U^t=1/c_{\rm grav}$.  Note, the mass densities and the pressures 
in Eqs. (\ref{matem2}) and (\ref{phem2}) are different from those in 
Eqs. (\ref{emmat}) and (\ref{bistens}), although they are denoted with the 
same notations.  In the comoving frame for the matter metric, the 
Friedmann equations therefore take the forms
\begin{equation}
\left({\dot{a}\over a}\right)^2={{8\pi G}\over 3}\rho+{c^2_{\rm grav}\over 
3}\Lambda-{{kc^2_{\rm grav}}\over a^2},
\label{frd21}
\end{equation}
\begin{equation}
{\ddot{a}\over a}-{\dot{c}_{\rm grav}\over c_{\rm grav}}{\dot{a}\over a}=
-{{4\pi G}\over 3}\left(\rho+3{p\over c^2_{\rm grav}}\right)+
{c^2_{\rm grav}\over 3}\Lambda,
\label{frd22}
\end{equation}
where
\begin{equation}
\rho\equiv\rho_{\Phi}+\hat{\rho}{c\over c_{\rm grav}},\ \ \ \ \ \ \ \ \ \ \ \ 
p\equiv p_{\Phi}+\hat{p}{c\over c_{\rm grav}}.
\label{mssprs2}
\end{equation}
Note, although we are now considering the comoving frame for the matter 
metric, the total energy momentum tensor ${\cal T}^{\mu\nu}$ is still 
defined w.r.t. the gravity metric.  It is just that the time coordinate 
$\tau$ in the gravity metric is the comoving frame time coordinate for 
the matter metric.  So, mass density and pressure of matter fields still 
have the factor of $c/c_{\rm grav}$ in Eq. (\ref{mssprs2}).  
From these Friedmann equations, we obtain the following conservation equation
\begin{equation}
\dot{\rho}+3\left(\rho+{p\over c^2_{\rm grav}}\right){\dot{a}\over a}=
{3\over{4\pi G}}{\dot{c}_{\rm grav}\over c_{\rm grav}}\left({\dot{a}\over a}
\right)^2-{{c_{\rm grav}\dot{c}_{\rm grav}}\over{4\pi G}}\Lambda+
{3\over{4\pi G}}{{kc_{\rm grav}\dot{c}_{\rm grav}}\over a^2}.
\label{conseq2}
\end{equation}

We now study the compatibility of the bimetric VSL cosmology with the second 
law of thermodynamics.  Although we are now considering the comoving frame for 
the matter metric, we first study thermodynamics in the frame of the gravity 
metric, since the conservation equation (\ref{conseq2}) is expressed in the 
gravity metric frame.  The first law of thermodynamics, applied to the 
comoving volume element of unit coordinate volume and physical volume $v=a^3$, 
takes the form
\begin{equation}
Tds=d(\rho c^2_{\rm grav}v)+pdv.
\label{1stlaw22}
\end{equation}
From Eqs. (\ref{conseq2}) and (\ref{1stlaw22}), we obtain
\begin{equation}
T\dot{s}=4\rho a^3c_{\rm grav}\dot{c}_{\rm grav},
\label{dots}
\end{equation}
where we made use of Eq. (\ref{frd21}) to simplify the RHS.  So, the total 
entropy $S=s\int dx^3\sqrt{-\gamma}$ increases with time while $c_{\rm grav}
=c/\sqrt{I}$ increases to the present day value.  The second law of 
thermodynamics is therefore always satisfied for any values of $k$.  
Next, we consider the second law of thermodynamics in the frame of the matter 
metric.  The mass density and the pressure in the matter metric frame are 
given by $\rho c_{\rm grav}/c$ and $p c_{\rm grav}/c$, and the energy density 
is obtained by multiplying the mass density by $c^2$.  So, the first law of 
thermodynamics takes the form:
\begin{equation}
Tds=d(\rho cc_{\rm grav}v)+{c_{\rm grav}\over c}pdv.
\label{1stlawm2}
\end{equation}
This along with Eq. (\ref{conseq2}) leads to
\begin{equation}
\dot{s}=3\left({{c^2_{\rm grav}-c^2}\over{c^2c^2_{\rm grav}}}p{\dot{a}\over a}
+\rho{\dot{c}_{\rm grav}\over c_{\rm grav}}\right){{cc_{\rm grav}a^3}\over T}.
\label{dotsm2}
\end{equation}
So, the terms in the parenthesis of this equation has to be non-negative in 
order for the the condition for the second law of thermodynamics to be 
satisfied.  Making use of the explicit expressions for $c_{\rm grav}$, $\rho$ 
and $p$, one can put the condition in the form: 
\begin{equation}
\dot{\Phi}^2\left[{1\over 2}{I\over c^2}\dot{\Phi}^2-V+\hat{p}\sqrt{I}
\right]{\dot{a}\over a}+\dot{\Phi}\ddot{\Phi}\left[{1\over 2}{I\over c^2}
\dot{\Phi}^2+V+\hat{\rho}c^2\sqrt{I}\right]\leq 0.
\label{2ndlawcnd}
\end{equation}
Making use of the equation of motion for the biscalar field $\Phi$, it can 
be shown that regardless of value of $k$ this condition can be satisfied by 
the perfect fluid obeying the equation of state $\hat{p}=(\gamma-1)\hat{\rho}
c^2$ with $\gamma\leq 2$, provided the biscalar potential satisfies 
$\partial_{\tau}V(\Phi)>0$. 

We discuss thermodynamics associated with matter fields, only.  As before, the 
equations of motion for the matter fields imply the conservation law 
(\ref{matencons}) for the matter fields energy-momentum tensor.  Since the 
speed of a photon takes a constant value $c_{\rm ph}=c$ in the comoving frame 
for the matter metric, the $t$-component of the matter fields conservation 
equation takes the form:
\begin{equation}
\dot{\hat{\rho}}+3\left(\hat{\rho}+{\hat{p}\over c^2}\right)
{\dot{a}\over a}=0.
\label{conseqbs2}
\end{equation}
Unlike the case of the comoving frame for the gravity metric, the matter 
fields are observed to be conserved.  Eq. (\ref{conseqbs2}) along with the 
first law of thermodynamics for the matter fields
\begin{equation}
Tds_{\rm mat}=d(\hat{\rho}c^2v)+\hat{p}dv,
\label{matf1st}
\end{equation}
yields $\dot{s}_{\rm mat}=0$.  So, time variation of entropy density $s$ 
in the matter metric frame, as expressed in Eq. (\ref{dotsm2}), is all due 
to time variation of entropy density of the biscalar field, and there is no 
entropy exchange between matter and the biscalar field sectors.  

\section{Conclusions}

We studied the compatibility of the VSL cosmological models with the second 
law of thermodynamics.  We find that generally the VSL model with varying 
fundamental constant is severely constrained by the second law of 
thermodynamics.  For the model with the speed of light gradually varying as 
a power law of cosmic scale factor, only open universe with positive 
cosmological constant is allowed to contain cosmic perfect fluid satisfying 
reasonable equation of state such as dust and radiation.  On the other hand, 
the bimetric cosmological models are less constrained by the second law of 
thermodynamics.  In the comoving frame associated with the gravity metric, 
there is no constraint on the model by the second law of thermodynamics.  
In the comoving frame associated with the matter metric, provided the 
biscalar potential satisfies a certain condition, regardless of the value 
of $k$, the second law of thermodynamics can be satisfied by perfect fluid 
satisfying the equation of state $\hat{p}=(\gamma-1)\hat{\rho}c^2$ with 
$\gamma\leq 2$, which includes dust and radiation.

\end{document}